\documentclass[aps, prb, amsmath,amssymb, reprint, nofootinbib,superscriptaddress]{revtex4-2}

\usepackage{orcidlink}

\usepackage{graphicx}
\usepackage{bm}

\usepackage{hyperref}

\usepackage{etoolbox}

\begin{document}

\title{Three-photon photoemission on metallic transition metal dichalcogenide NiTe$_{2}$}

\author{Cheng-Tien Chiang \orcidlink{0000-0003-4605-3207}$^{\diamond}$}
\email{ctchiang@as.edu.tw}
\homepage{http://ufss.iams.sinica.edu.tw/}
\affiliation{Institute of Atomic and Molecular Sciences, Academia Sinica, Taipei, Taiwan}
\affiliation{Molecular Science and Technology Program, Taiwan International Graduate Program, Academia Sinica, Taiwan}
\affiliation{Department of Physics, National Taiwan University, Taipei, Taiwan}
\affiliation{Department of Physics, National Sun Yat-sen University, Kaohsiung, Taiwan}
\affiliation{Department of Physics, National Central University, Taoyuan, Taiwan}

\author{Mukesh Singh \orcidlink{0009-0002-5092-5266}$^{\diamond}$}
\affiliation{Institute of Atomic and Molecular Sciences, Academia Sinica, Taipei, Taiwan}
\affiliation{Department of Physics, Indian Institute of Science Education and Research (IISER) Mohali, Punjab, India}

\author{\mbox{Yu-Jhen Chuang \orcidlink{0009-0001-3850-3098}}}
\affiliation{Department of Physics, National Taiwan University, Taipei, Taiwan}
\affiliation{Institute of Atomic and Molecular Sciences, Academia Sinica, Taipei, Taiwan}

\author{\mbox{Chia-Nung Kuo \orcidlink{0009-0008-1382-0610}}}
\affiliation{Academy of Innovative Semiconductor and Sustainable Manufacturing, National Cheng Kung University, Tainan 70101, Taiwan}
\affiliation{Department of Physics, National Cheng Kung University, Tainan 70101, Taiwan}
\affiliation{Taiwan Consortium of Emergent Crystalline Materials (TCECM), National Science and Technology Council, Taipei 106214, Taiwan}

\author{\mbox{Prabesh Bista \orcidlink{0000-0002-6910-4620}}}
\affiliation{Department of Physics, National Central University, Taoyuan, Taiwan}
\affiliation{Molecular Science and Technology Program, Taiwan International Graduate Program, Academia Sinica, Taiwan}
\affiliation{Institute of Atomic and Molecular Sciences, Academia Sinica, Taipei, Taiwan}
\affiliation{Department of Physics, Stony Brook University, New York, USA}

\author{\mbox{Yu-Chan Lin \orcidlink{0000-0001-5405-6315}}}
\affiliation{Institute of Atomic and Molecular Sciences, Academia Sinica, Taipei, Taiwan}
\affiliation{National Center for Instrumentation Research, National Institutes of Applied Research, Hsinchu, Taiwan}

\author{\mbox{Chen-Bin Huang \orcidlink{0000-0002-5824-3318}}}
\affiliation{\mbox{Institute of Photonics Technologies, National Tsing Hua University, Hsinchu, Taiwan}}
\affiliation{Research Center for Applied Sciences, Academia Sinica, Taipei, Taiwan}

\author{\mbox{Ming-Chiang Chung \orcidlink{0000-0002-8538-0422}}}
\affiliation{Department of Physics, National Chung Hsing University, Taichung, Taiwan}
\affiliation{Physics Division, National Center for Theoretical Sciences, Taipei, Taiwan}
\affiliation{Donostia International Physics Center (DIPC), San Sebastián, Spain}

\author{\mbox{Chin Shan Lue \orcidlink{0000-0002-3074-9253}}}
\affiliation{Academy of Innovative Semiconductor and Sustainable Manufacturing, National Cheng Kung University, Tainan 70101, Taiwan}
\affiliation{Department of Physics, National Cheng Kung University, Tainan 70101, Taiwan}
\affiliation{Taiwan Consortium of Emergent Crystalline Materials (TCECM), National Science and Technology Council, Taipei 106214, Taiwan}

\date{\today}

\begin{abstract}
Three-photon photoemission (3PPE) signals on NiTe$_{2}$ are observed for the first time upon ultraviolet excitations. In the energy range beyond two-photon excitations, an energy-momentum dispersive feature is clearly identified with characteristic light polarization dependence. Based on its dispersion, spectral width, as well as intensity as a function of the parallel momentum, this dispersive 3PPE feature could be ascribed to an image potential resonance overlapping with an unoccupied Ni $4s$\,-\,Te $4d/5p$ bulk band in accordance with the electronic band structure. Our results underlie higher-order non-linear photoemission on metallic transition metal dichalcogenides and facilitate future ultrafast studies of coherent electron dynamics in anisotropic, two-dimensional systems.
\end{abstract}

\maketitle

\def\thefootnote{${\diamond}$}\footnotetext{These authors contributed equally to this work.}\def\thefootnote{\arabic{footnote}}

\section{Introduction}

Non-linear photoemission has become a versatile tool to study high-order light-matter interaction at solid surfaces \cite{Reutzel_PRX19,Li_PRB22,Aeschlimann_SurfSci25,Yen_PRX26}. Beginning with its lowest order, two-photon photoemission (2PPE) has revealed coherent evolution of transiently excited electron wave function on the femtosecond scale \cite{Guedde_Science07,Marks_PRB11}. At higher orders, not only the electronic band structure plays an even more important role due to the increasing possibilities of multi-photon resonances \cite{Li_NJP20,Winkelmann_PRB09}, but also the signatures of collective optical responses could be identified \cite{Reutzel_PRL19,Shen_NatComm25,Cui_NatPhys14}. In contrast to non-linear optics, optically excited electrons are directly detected in non-linear photoemission, providing straightforward access to the energy and momentum phase space of the non-linear transitions \cite{Bisio_PRL06,Li_NJP20}.

As a significant extension of non-linear photoemission on conventional metals \cite{Petek_PSS97,Weinelt_JPCM02,Ferrini_NIMPRA09,Fitzgerald_PRB13,Sirotti_PRB14,Tan_PRX17,Reutzel_PRB20,Dreher_CommPhys23,Tai_ApplPhysExp25}, semiconductors \cite{Bloch_PRL96,Mihaychuk_PRB99,Wiets_PRB03,Eickhoff_PRL11,Marsell_NanoLett17}, and graphene \cite{Armbrust_PRL12,Takahashi_PRB14,Niesner_JPCM14,Gugel_2Dmater15,Montagnese_SciRep16,Tognolini_SurfSci19}, there are a few pioneering 2PPE studies on transition metal dichalcogenides \cite{Eul_ACSPhoton25,Dai_NanoLett21,Xu_ACSNano24,Wiesenmayer_PRB10,Tanaka_PRB03}, h-BN \cite{Muntwiler_PRB07,Fukumoto_JPD20,Li_NatComm23}, and black phosphorus \cite{Joshi_NanoLett22,Shen_NatComm25} with two-dimensional (2D) layered structure. Only very recently the higher-order three- and four-photon photoemission (3-/4PPE) have been observed on the semiconducting MoS$_{2}$, SnSe$_{2}$ \cite{Liu_PCCP21,Jiao_FrontPhys24}, black phosphorus \cite{Joshi_NanoLett22}, and h-BN \cite{Hengsberger_JPD20} using \textit{infrared} and \textit{visible light} excitations. In this work we report on the first observation of 3PPE from the metallic 2D material NiTe$_{2}$ upon \textit{ultraviolet} excitations. These 3PPE signals consist of a clear energy-momentum dispersive feature on top of an exponentially decaying distribution towards higher energies. While the latter persist for both incident light polarization directions, the former can be clearly identified with $p$-polarized excitation. The nature of this dispersive, polarization dependent 3PPE feature is discussed with the electronic band structure of NiTe$_{2}$ \cite{Fischer_PRB24,Rizza_ACSApplNanoMater22,Ferreira_PRB21,Manesco_Zendo20,Guo_JPC86}, and it can be attributed to an image potential resonance hybridized with a bulk unoccupied Ni $4s$\,-\,Te $4d/5p$ band. Our results represent the first step towards non-linear photoemission on metallic transition metal dichalcogenides, allowing future time-resolved studies of their anisotropic, optically excited electronic structure \cite{Sharma_PRB25,Rizza_ACSApplNanoMater22}.

\begin{figure}[t!]
\center{\includegraphics[width=1\columnwidth]{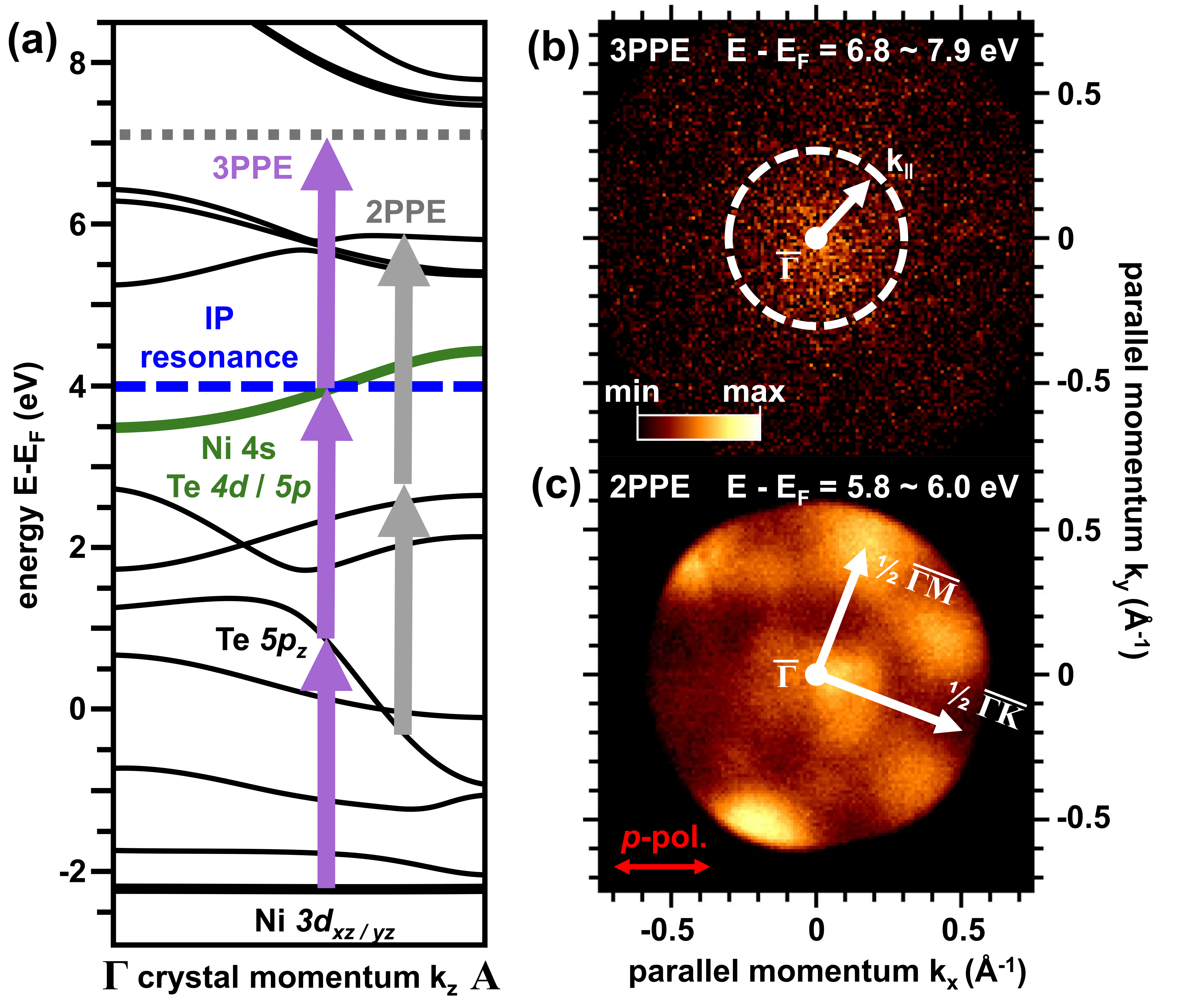}}
\caption{(a) Three- and two-photon photoemission (3PPE,\,2PPE) on NiTe$_2$ with the photon energy $h\nu$ around 3.1\,eV according to theoretical electronic band structure \cite{Fischer_PRB24,Ferreira_PRB21,Manesco_Zendo20}. The type II Dirac point is fixed at the Fermi energy ($E_{F}$), and the bands from 1.5\,eV below $E_{F}$ up to 3\,eV above $E_{F}$ are adapted from the {GW} calculations \cite{Fischer_PRB24}. The occupied Ni $3d$ bands around 2\,eV below $E_{F}$, the unoccupied Ni $4s$ band hybridized with the Te $4d$ and $5p$ orbitals near 4\,eV above $E_{F}$ (green) \cite{Rizza_ACSApplNanoMater22,Guo_JPC86}, as well as the higher unoccupied bands are from Refs.\,\cite{Ferreira_PRB21,Manesco_Zendo20}. The bands in the range of 5 to 6.5\,eV above $E_{F}$ are shifted upwards by about 0.1\,eV in accordance with our recent 2PPE results \cite{Singh_APL26}. Blue dashed horizonal line indicates the possible image potential (IP) resonance, which will be discussed in Sec.\,\ref{sec:discussion} together with the Ni $4s$\,-\,Te $4d/5p$ band as the relevant intermediate states for the 3PPE feature at the final state energy near 7\,eV above $E_{F}$ (gray dotted). The type I Dirac point near 2\,eV above $E_{F}$ as well as the type II one could be sensitive to strain and many-body corrections \cite{Ferreira_PRB21,Fischer_PRB24}. (b,c) Momentum distribution of photoelectrons in 3PPE and 2PPE experiments with $p$-polarized light at $h\nu$\,=\,3.09\,$\pm$0.03\,eV, with the former integrated over about 6.8 to 7.9\,eV above $E_{F}$, and the latter from 5.8 to 6.0\,eV above $E_{F}$. In (b) the radial parallel momentum $k_{\|}$\,=\,$\sqrt{k_{x}^{2}+k_{y}^{2}}$ for Figs.\,\ref{fig3}-\ref{fig5} is illustrated, with $k_{x,y}$ as the photoelectron momentum components parallel to the surface. High symmetry directions \cite{Singh_APL26} and the electric field direction of the incident $p$-polarized light are shown in (c). The (minimum, maximum) values of the color scale are about (1,\,25) and (2.8$\times10^{2}$,\,1.6$\times10^{3}$) counts for (b) and (c), respectively.}
\label{fig1}
\end{figure}

\begin{figure}[t!]
\center{\includegraphics[width=1\columnwidth]{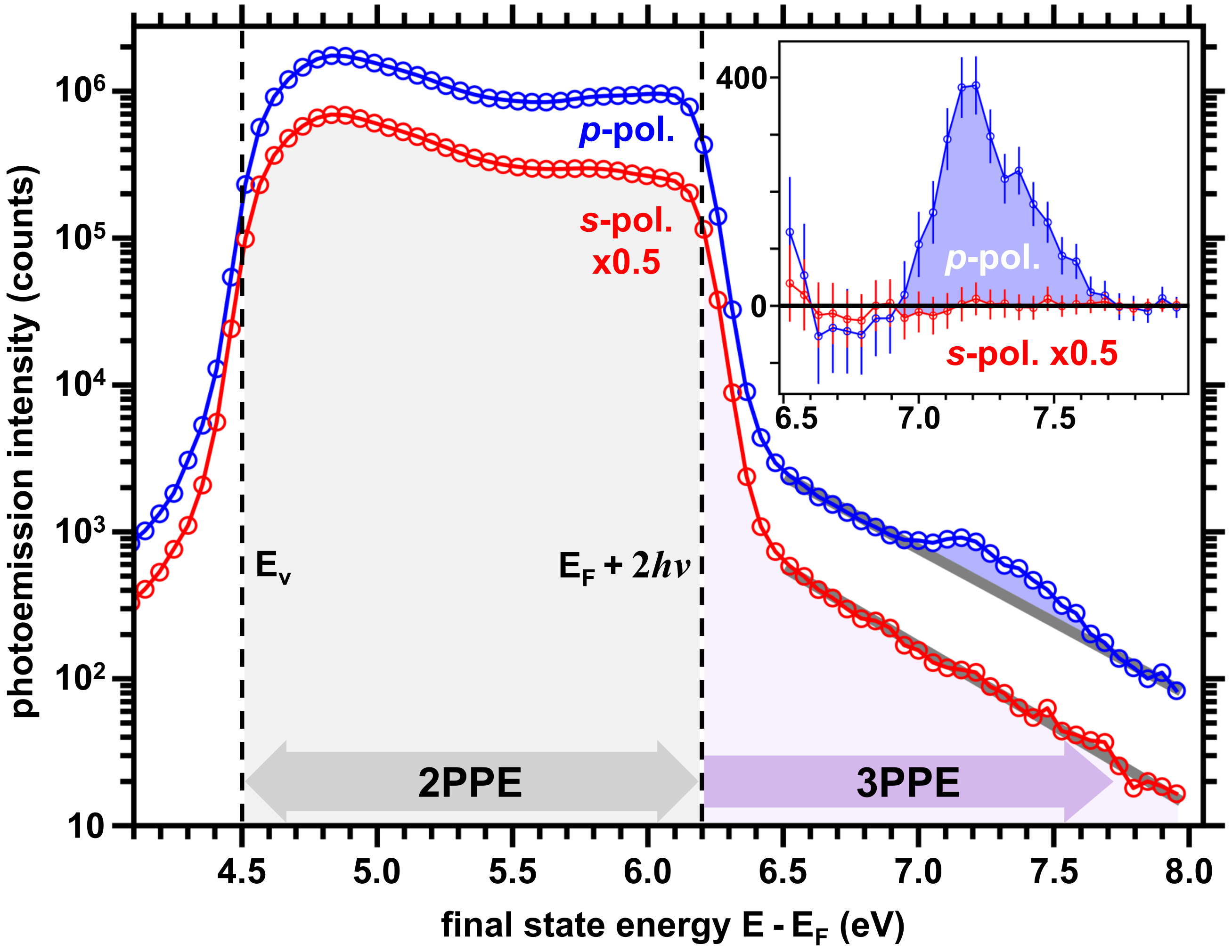}}
\caption{Momentum-integrated 2PPE and 3PPE spectra measured with $p$- and $s$-polarized light at the photon energy of $h\nu$\,=\,3.09\,$\pm$0.03\,eV. 2PPE extends over the energy range from the work function cutoff ($E_{v}$\,\cite{Ghosh_PRB19,Di_APL23,Je_EleMaterLett23}) to the feature of 2PPE from the Fermi level ($E_{F}$\,+\,2$h\nu$), above which 3PPE signal becomes visible \cite{Ferrini_PRL04,Banfi_PRL05,Bisio_PRL06,Schattke_PRB08,Hao_PRL10,Bisio_JPCM11,Reutzel_PRX19,Reutzel_PRB20,Li_PRB22}. The 3PPE spectra acquired with the $s$-polarized light can be described by an exponential decay towards higher energies with a decay constant of 400\,$\pm$\,40\,meV (gray) \cite{Banfi_PRB03}, whereas that for the $p$-polarized light consists of a similar decay with 440\,$\pm$\,40\,meV (gray) and an additional feature (blue filled). This feature can be more clearly seen in the inset on the linear intensity scale, where the exponential backgrounds are subtracted. For clarity, the $s$-polarized spectra are displayed with a factor of 0.5.}
\label{fig2}
\end{figure}

\section{Experiment and results}

Single crystals NiTe$_{2}$ were cleaved in the ultrahigh vacuum chamber, and their growth have been described previously \cite{Ghosh_PRB19,Nappini_AdvFunMat20,Zhang_NatComm21,Rizza_ACSApplNanoMater22}. The photoelectrons are excited by the frequency-doubled output of a home-built non-collinear optical amplifier, which has been constructed according to the design by Wittmann \textit{et al.} \cite{Wittmann19,Huber_RSI19}. The energy and momentum distributions of the photoelectrons are analyzed by the commercially available time-of-flight momentum microscope \cite{ToFMM,Schoenhense_JVSTA22,Schoenhense_ELSPEC15}, and all experiments were performed at room temperature. Details of our experiments can be found in a recent publication \cite{Singh_APL26}.

\subsection{Overview of 3PPE / 2PPE}

The relevant non-linear photoemission processes on NiTe$_{2}$ are indicated in the band structure in Fig.\,\ref{fig1}(a), and in Fig.\,\ref{fig1}(b,c) the momentum patterns of photoelectrons from 3PPE and 2PPE excited by $p$-polarized light are shown. The main intensity of the energy-integrated 3PPE signal in Fig.\,\ref{fig1}(b) is concentrated around the origin of the 2D momentum space, in strong contrast to the characteristic anisotropic 2PPE momentum distribution in Fig.\,\ref{fig1}(c) as previously reported \cite{Singh_APL26}. The momentum-integrated photoemission energy spectra are shown in Fig.\,\ref{fig2} for both $p$- and $s$-polarized light. There a clear 3PPE feature can be identified on the logarithmic intensity scale as indicated by the blue filled area when using the $p$-polarized light. This distinct 3PPE feature is located on top of an exponential background (gray), which is observed for both light polarization directions here and comparable to earlier non-linear photoemission on noble metals \cite{Banfi_PRB03,Bisio_PRL06,Sirotti_PRB14,Tan_PRX17}. In the inset of Fig.\,\ref{fig2} the 3PPE spectra after subtracting this exponential background are displayed on the linear scale, and the clear 3PPE feature can be seen. With the 3PPE momentum pattern in Fig.\,\ref{fig1}(b) and the polarization-dependent 3PPE spectra in Fig.\,\ref{fig2}, we can identify the relevant energy-momentum coordinates of the 3PPE feature in the range of 7.0 to 7.6\,eV above the Fermi level ($E_{F}$) near the $\overline{\Gamma}$ point.

\begin{figure}[t!]
\center{\includegraphics[width=1\columnwidth]{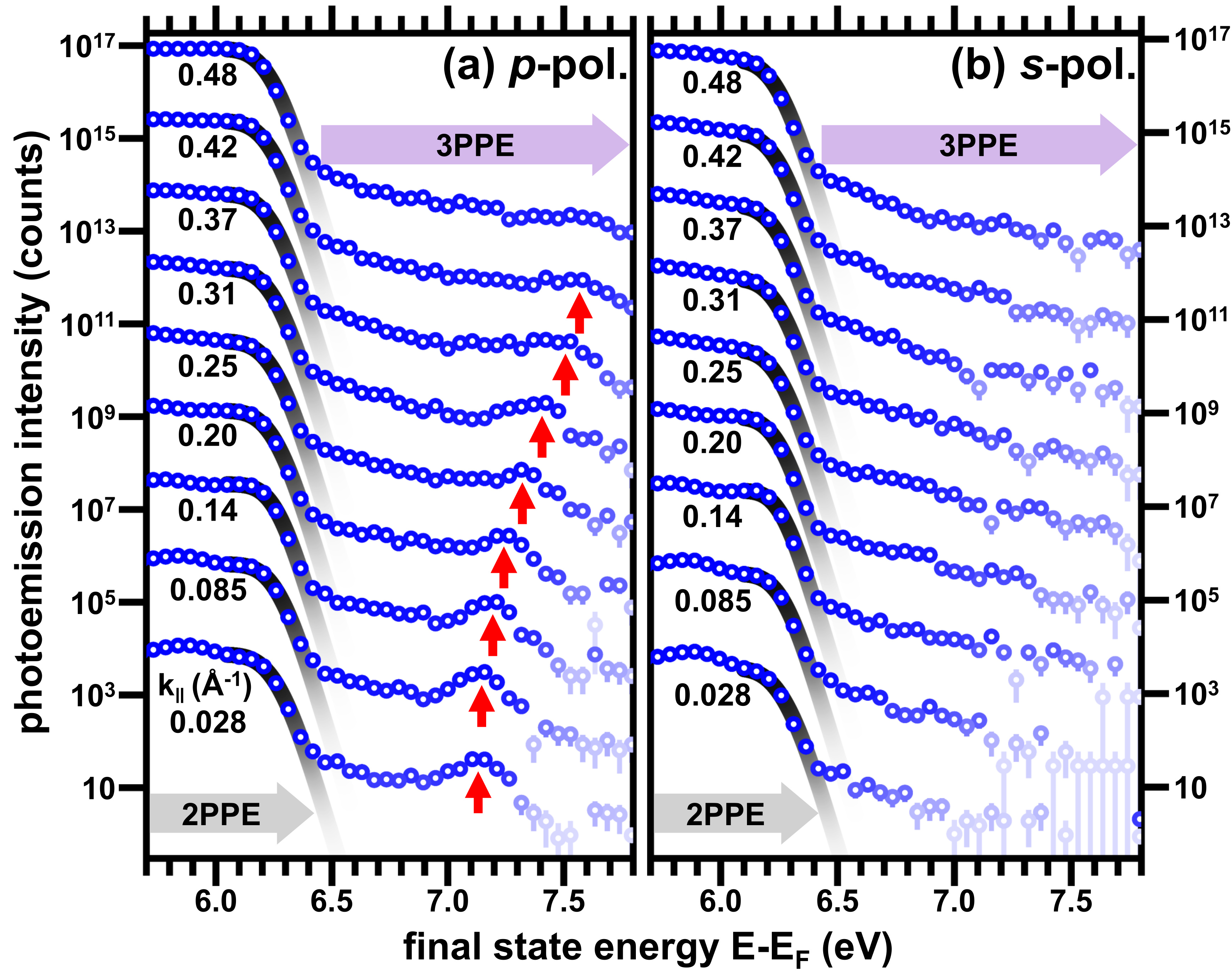}}
\caption{Momentum-dependent 2PPE and 3PPE spectra measured with (a) $p$- and (b) $s$-polarized light, each of which is integrated around the central parallel momentum $k_{\|}$ in \AA$^{-1}$ in Fig.\,\ref{fig1}(b) with an integration range of $\pm$\,0.028\,\AA$^{-1}$. 2PPE from the Fermi level ($E_{F}$) is indicated by the gray curves, describing Fermi-Dirac distributions at 300\,$\pm$\,10\,K convoluted by a Gaussian function with a full-width-at-half-maximum of 110\,$\pm$\,20\,meV. For clarity, each of the spectra in (a,b) from the second at the bottom to the top most one is multiplied by a factor of 30. The data points are color-coded in blue with their shade reciprocal to the intensity uncertainty, and the convoluted Fermi-Dirac distributions in gray scale according to their logarithmic intensity. Red vertical arrows indicate the dispersive 3PPE features, which are shown in detail on the linear scale in Fig.\,\ref{fig4}.}
\label{fig3}
\end{figure}

\begin{figure}[t!]
\center{\includegraphics[width=1\columnwidth]{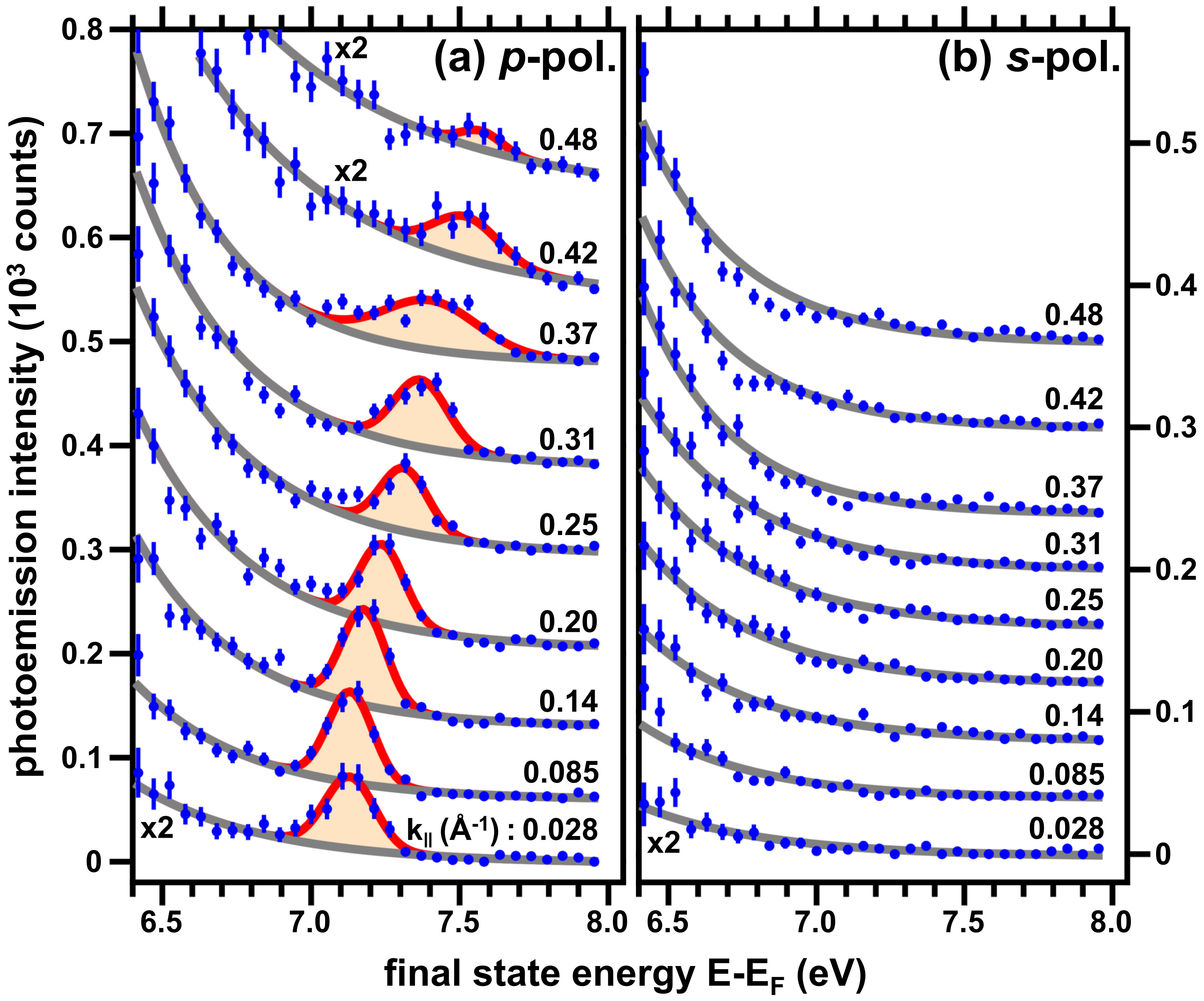}}
\caption{Momentum-dependent 3PPE spectra from Fig.\,\ref{fig3}(a) as a function of the parallel momentum $k_{\|}$ in \AA$^{-1}$, which is indicated in Fig.\,\ref{fig1}(b). (a) Spectra measured with incident $p$-polarized light, after subtracting the Fermi-Dirac distribution convoluted by the Gaussian function as shown in Fig.\,\ref{fig3}. Each background-subtracted spectrum can be described by a Gaussian function (red\,\cite{GaussFun}) on top of an exponentially decaying background (gray). The central energies, widths, as well as the peak areas of the Gaussian functions are shown in Fig.\,\ref{fig5}, whereas the decaying constants of the exponential functions from small to large $k_{\|}$ are 430, 380, 350, 370, 370, 350, 310, 510, and 580\,$\pm$\,30\,meV. For clarity, the bottom-most and the top two spectra are multiplied by a factor of two, and starting at the second from the bottom to large $k_{\|}$ they are shifted vertically by 0.06, 0.13, 0.21, 0.30, 0.38, 0.48, 0.54, and 0.64$\times$10$^{3}$\,counts. (b) Similar to (a) but acquired with $s$-polarized light, and each spectrum can be described by an exponentially decaying function with a decay constant of 430, 370, 380, 380, 370, 370, 300, 310, 320\,$\pm$\,50\,meV from small to large $k_{\|}$.  The bottom-most spectrum is multiplied by a factor of two, and from the second smallest $k_{\|}$ to the top one their vertical shifts are 0.04, 0.08, 0.12, 0.16, 0.20, 0.24, 0.30, and 0.36$\times$10$^{3}$\,counts.}
\label{fig4}
\end{figure}

\subsection{Momentum-dependent 3PPE}

\begin{figure}[t!]
\center{\includegraphics[width=0.8\columnwidth]{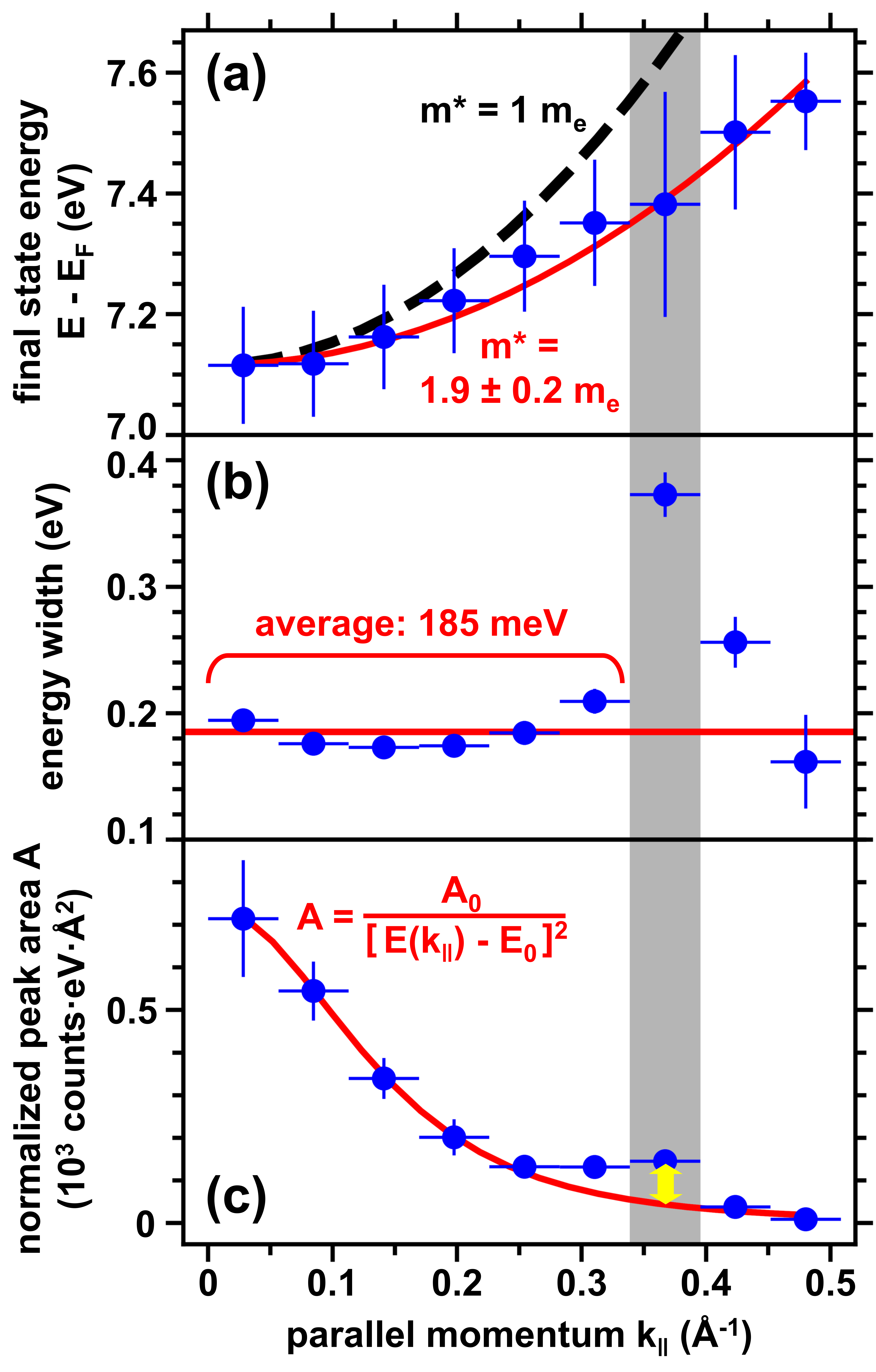}}
\caption{Dispersion, width, and intensity of the fitted 3PPE peak features in Fig.\,\ref{fig4}(a) as a function of the parallel momentum $k_{\|}$ in Fig.\,\ref{fig1}(b), with the $k_{\|}$ coordinates and their error bars indicating the center and width of the integration range, respectively. (a) Energy-momentum dispersion $E(k_{\|})$ relative to the Fermi energy ($E_{F}$), and the energy error bars indicate the full-width-at-half-maximum as displayed in (b). The red curve in (a) is a parabolic fit using an effective mass $m^{*}$\,$\approx$\,1.9\,$\pm$\,0.2\,$m_{e}$, and the dashed one with $m^{*}$\,=\,$m_{e}$ for comparison.  Here $m_{e}$ is the rest mass of electron, and the minimal of both curves $E(k_{\|}$\,=\,0$)$ are fixed at the measured minimal at 7.1\,$\pm$\,0.1\,eV. In (b) the average from $k_{\|}$\,=\,0.028 to 0.31\,\AA$^{-1}$ is indicated by the horizon line. (c) Peak area $A$ normalized to the integrated momentum area. The red curve is a phenomenological description $A$\,=\,$A_{0}/[E(k_{\|})-E_{0}]^{2}$, with $A_{0}$\,=\,6\,$\pm$\,1\,counts$\cdot$eV$^{3}\cdot$\AA$^{2}$ and $E_{0}$\,=\,7.0\,$\pm$\,0.1\,eV. The vertical gray area at around $k_{\|}$\,=\,0.37\,\AA$^{-1}$ indicates a significant broadening in (b) and slightly higher 3PPE intensity in (c) (yellow arrow).}
\label{fig5}
\end{figure}

To reveal the energy-momentum dispersion of the observed 3PPE feature under $p$-polarized excitation, in Fig.\,\ref{fig3} the momentum-dependent photoemission spectra are shown on the logarithmic scale. Here the azimuthal direction is integrated, and the radial momentum $k_{\|}$ dependence can be traced with sufficient statistics. As indicated by the arrows in Fig.\,\ref{fig3}(a), the 3PPE feature identified in the momentum-integrated spectrum in Fig.\,\ref{fig2} has a clear $k_{\|}$ dependence, which is only present with $p$-polarized excitation and absent with the $s$-polarized light in Fig.\,\ref{fig3}(b). A closer look at these $k_{\|}$-dependent 3PPE spectra on the linear scale in Fig.\,\ref{fig4}(a) indicates that the dispersive 3PPE feature can be reasonably described by a Gaussian distribution (red) \cite{GaussFun} sitting on top of an exponential background (gray), and only the latter is present when illuminated with $s$-polarized light in Fig.\,\ref{fig4}(b). The peak position, width, as well as area of these Gaussian functions are summarized in Fig.\,\ref{fig5} as a function of $k_{\|}$.

As displayed in Fig.\,\ref{fig5}(a), the peak position of the 3PPE feature in Fig.\,\ref{fig4}(a) disperses monotonically towards higher energies as the momentum $k_{\|}$ increases (blue circle). This dispersion can be described by the parabolic dispersion using an effective mass of 1.9\,$\pm$\,0.2\,$m_{e}$ (red curve), with $m_{e}$ as the electron mass. For comparison, the free electron dispersion is shown by the dashed curve, which could only match the observed dispersion in the smaller momentum region of $k_{\|}$\,$\lesssim$\,0.3\,\AA$^{-1}$. In Fig.\,\ref{fig5}(b) the full-width-at-half-maximum of the 3PPE feature is depicted, which varies only slowly for $k_{\|}$\,$\lesssim$\,0.3\,\AA$^{-1}$ with an abrupt increase at $k_{\|}$\,$\approx$\,0.37\,\AA$^{-1}$ (gray area) and decreases afterwards. The peak area $A$ of the 3PPE feature, normalized by the $k_{\|}$-dependent area in the momentum space, is shown in Fig.\,\ref{fig5}(c) and decreases monotonically towards higher $k_{\|}$ within $k_{\|}$\,$\lesssim$\,0.2\,\AA$^{-1}$, consistent with the energy-integrated 3PPE momentum distribution in Fig.\,\ref{fig1}(b). At much higher $k_{\|}$, $A$ stays roughly constant until $k_{\|}$\,$\approx$\,0.37\,\AA$^{-1}$ and decreases thereafter. In the following section, the nature of this dispersive 3PPE feature will be discussed in accordance with the observation in Fig.\,\ref{fig5}.

\section{Discussion\label{sec:discussion}}

As summarized in Fig.\,\ref{fig5}(a), the dispersive 3PPE feature has a minimum final state energy of $E$\,-\,$E_F$\,=\,7.1\,$\pm$\,0.1\,eV at $k_{\|}$\,$\approx$\,0, which can be translated into an intermediate state at a binding energy of 0.5\,$\pm$\,0.2\,eV below the vacuum level at 4.5\,$\pm$\,0.1\,eV ($E_{v}$ in Fig.\,\ref{fig2}) \cite{Ghosh_PRB19,Di_APL23,Je_EleMaterLett23}. This binding energy is compatible with the $n$\,=\,1 image potential (IP) state on TiS$_{2}$ at 0.60\,$\pm$0.1\,eV \cite{Straub_PRB86}, on TiTe$_{2}$ at 0.7\,$\pm$0.3\,eV \cite{Drube_JPC87}, on GeTe at 0.72\,$\pm$0.01\,eV \cite{Chassot_arXiv25}, on black phosphorus at 0.72\,$\pm$0.06\,eV \cite{Shen_NatComm25}, on Ni(110) and (111) at 0.6\,$\pm$0.3\,eV \cite{Goldmann_PRB85}, as well as the IP resonance on Cu(110) at 0.7\,$\pm$0.1\,eV \cite{Quiniou_PRB93}. For comparison, the recent scanning tunneling spectroscopy by Fang \textit{et al}. observed the first field emission resonance of an NiTe$_{2}$ film at around 4.8\,eV above $E_{F}$ with a broad full-width-at-half-maximum of about 0.8\,eV \cite{Fang_ChinPhysB26}, and it could be related to the energy-shifted $n$\,=\,1 IP state with the tip-induced Stark effect \cite{Binning_PRL85,Wahl_PRL03,Dougherty_PRB07}.

Since the binding energy of the observed 3PPE feature at $k_{\|}$\,$\approx$\,0 in Fig.\,\ref{fig5}(a) is compatible with the $n$\,=\,1 IP state, its energy-momentum dispersion needs to be considered. According to Fig.\,\ref{fig5}(a), the dispersion can be fairly described by an effective mass $m^{*}$\,$\approx$\,1.9\,$m_{e}$ over the whole investigated momentum $k_{\|}$ range. This mass is much heavier than the $n$\,=\,1 IP state on MoS$_{2}$ with 1.05\,$\pm$0.06\,$m_{e}$  \cite{Liu_PCCP21}, on VSe$_{2}$ with 1.3\,$\pm$0.2\,$m_{e}$ \cite{Claessen_JPCM90}, on TiTe$_{2}$ with around 1.5\,$m_{e}$ \cite{Drube_JPC87}, as well as on noble metals, graphite, and graphene within the range of 0.8 to 1.4\,$m_{e}$ \cite{Giesen_PRB86,Straub_PRB86,Yilmaz_JVSTA12,Montagnese_SciRep16,Niesner_JPCM14}. However, our $m^{*}$ is comparable to the IP resonance on Cu(110) with 1.8 to 1.9\,$\pm$0.2\,$m_{e}$ \cite{Sonoda_PRB11}, as well as the IP state on Cu(111) with 1.6 to 2.2\,$\pm$0.1\,$m_{e}$ \cite{Pagliara_SurfSci08}, on Ni(110) and (111) with 1.6 to 1.7\,$\pm$0.3\,$m_{e}$ \cite{Goldmann_PRB85}, on the ferroelectric MoTe$_{2}$ with 1.6\,$\pm$0.2\,$m_{e}$ \cite{Dai_NanoLett21}, and on the rippled graphene supported by Ru(0001) with 2.1\,$\pm$0.8\,$m_{e}$ \cite{Armbrust_PRL12}. In view of the large $m^{*}$ in Fig.\,\ref{fig5}(a) and the presence of a bulk band near 4\,eV above the $E_{F}$ (green curve in Fig.\,\ref{fig1}(a)) near the intermediate state energy of the 3PPE feature, we tentatively assign the dispersive 3PPE feature to an IP resonance. This assignment would be consistent with the abrupt broadening at $k_{\|}$\,$\approx$\,0.37\,\AA$^{-1}$ in Fig.\,\ref{fig5}(b) due to the possible hybridization with the bulk band (gray), as well as the additional 3PPE intensity contribution there in Fig.\,\ref{fig5}(c) (arrow) on top of the decreasing 3PPE intensity towards larger $k_{\|}$ (red curve). Similar broadening has been observed by 2PPE and 3PPE via IP resonances on Cu and Ag surfaces due to overlap with bulk bands \cite{Li_PRB22,Damm_PRB09}.

Our assignment of the dispersive 3PPE feature to the IP resonance is further consistent with the light polarization dependence in Figs.\,\ref{fig2}-\ref{fig4}. As shown in Figs.\,\ref{fig2}-\ref{fig4}, the IP resonance is observed only when $p$-polarized light is applied. This can be ascribed to its wave function with a dominant hydrogenic part along the surface normal \cite{Tsirkin_PRB13,Marks_PRB11}, which is similar to the IP state and is fully symmetric upon rotation around the surface normal \cite{Echenique_JPC78,Gies_JPC86,Chulkov_SurfSci99,Echenique_ELSPEC02}. Assuming a photoemission final state wave function that is also totally symmetric around the surface normal, only the electric field component perpendicular to the surface can probe the IP resonance due to optical dipole selection rules \cite{Eberhardt_PRB80,Murray_JPC72,Mattheiss_PRB73,Fong_PRB73}. Since this electric field component is included in the $p$- but not in the $s$-polarization, the IP resonance can only be observed in the former case, in agreement with previous observations of IP states \cite{Banfi_PRL05,Bisio_PRL06,Pagliara_PRB13,Jiao_FrontPhys24}.

Last but not least, we discuss the role of the IP resonance assigned above in the 3PPE process in the electronic band structure. As schematically shown by the band structure along the surface normal in Fig.\,\ref{fig1}(a), the IP resonance can be populated via two-photon transition from the occupied Ni $3d_{xz,yz}$ bands via the intermediate Te $5p_{z}$ unoccupied band \cite{Fischer_PRB24,Ferreira_PRB21,Manesco_Zendo20,Rizza_ACSApplNanoMater22,Guo_JPC86}. The Ni $3d_{xz,yz}$ to Te $5p_{z}$ transition here is allowed for both the $p$- and $s$-polarization of light because the former belongs to the $\Delta_{3}$ symmetry and the latter the $\Delta_{1}$ \cite{Guo_JPC86,Murray_JPC72,Mattheiss_PRB73}. However, the second transition from the Te $5p_{z}$ band to the IP resonance is only allowed for the $p$-polarization because both these states belong to the $\Delta_{1}$ symmetry. This second transition together with the third photoexcitation transition into the final state, as discussed in the previous paragraph, lead to the overall polarization dependence. Moreover, since the Te $5p_{z}$ band disperses downwards in energy as the momentum parallel to the surface increases, the resonant condition in Fig.\,\ref{fig1}(a) would no longer be available at higher $k_{\|}$, leading to a decreasing 3PPE intensity towards higher $k_{\|}$ as summarized in Fig.\,\ref{fig5}(c).

\section{Summary}

In this work we report on the first observation of three-photon photoemission (3PPE) on NiTe$_{2}$. By using femtosecond ultraviolet excitations at a photon energy of 3.09\,$\pm$\,0.03\,eV and examining the polarization dependent photoemission intensity at higher energies beyond the two-photon photoemission, we clearly identify a 3PPE spectral feature at about 7.1\,$\pm$\,0.1\,eV above the Fermi energy ($E_{F}$). This 3PPE feature disperses upwards in energy as the momentum component parallel to the surface increases and can be described by an effective mass of 1.9\,$\pm$\,0.2\,m$_{e}$. Based on the polarization dependence, spectral width, and intensity in accompany with the available theoretical electronic band structure \cite{Fischer_PRB24,Ferreira_PRB21,Manesco_Zendo20,Rizza_ACSApplNanoMater22,Guo_JPC86} as well as symmetry properties of the bands \cite{Murray_JPC72,Mattheiss_PRB73}, we assign the dispersive signals to 3PPE via an intermediate, unoccupied image potential resonance centered at about 4.0\,\,eV above $E_{F}$. Our results benchmark higher-order non-linear photoemission processes on metallic transition metal dichalcogenides, which would allow future time-resolved studies of resonantly excited electron dynamics \cite{Reutzel_PRX19,Reutzel_PRL19} as well as ultrafast coherent control of electronic states over their 2D momentum space \cite{Guedde_Science07,Winkelmann_PRB09}.
\hfill \break

\acknowledgments
The authors thank financial support from the National Science and Technology Council (NSTC\,115-2112-M-001-047, NSTC\,115-2124-M-008-008, NSTC\,114-2112-M-001-050, NSTC\,112-2923-M-001-004-MY3, MOST\,111-2112-M-001-054-MY3, NSTC\,113-2112-M-006-009-MY2, MOST\,109-2112-M-007-031-MY3, NSTC\,115-2112-M-005-005), the Academia Sinica (AS-TP-111-M03), as well as the Asian Office of Aerospace Research and Development (AOARD) for support under Award No.\,FA2386-23-1-4104, and have no conflict to disclose.

\bibliography{ref}

\end{document}